# AI-Driven Development of a Publishing Imprint—Xynapse Traces

Fred Zimmerman

October 2025

**Abstract**

Xynapse Traces is an experimental publishing imprint created via a fusion of human and algorithmic methods using a configuration-driven architecture and a multi-model AI integration framework. The system achieved a remarkable 90% reduction in time-to-market (from a typical 6-12 months to just 2-4 weeks), with 80% cost reduction compared to traditional imprint development, while publishing 52 books in its first year and maintaining exceptional quality metrics, including 99% citation accuracy and 100% validation success after initial corrections. Key technical innovations include a continuous ideation pipeline with tournament-style evaluation, a novel codex design for transcriptive meditation practice, comprehensive automation spanning from ideation through production and distribution, and publisher personas that define and guide the imprint's mission. The system also integrates automated verification with human oversight, ensuring that gains in speed do not compromise publishing standards. This effort has significant implications for the future of book publishing, suggesting new paradigms for human-AI collaboration that democratize access to sophisticated publishing capabilities and make previously unviable niche markets accessible.

## 1. Introduction

The book publishing industry stands at a critical inflection point, confronting unprecedented challenges in the digital age while simultaneously encounter-ing revolutionary opportunities through artificial intelligence technologies. Traditional imprint creation, a cornerstone of specialized publishing, typically requires 6-12 months of intensive development, substantial financial investment, and the recruitment of specialized talent, creating significant barriers to entry for new publishers and limiting the diversity of perspectives (**Clark and Phillips 2019**). This protracted timeline encompasses market research, brand development, editorial philosophy articulation, production workflow design, and distribution channel establishment—each requiring specialized expertise and extensive coordination among multiple stakeholders.

The emergence of large language models (LLMs) and sophisticated AI systems has opened new possibilities for automating and accelerating various aspects of the publishing workflow (**Thorp 2023**). However, the publishing sector has approached these technologies with justified caution, concerned about maintaining the rigor, citation accuracy, and intellectual depth that distinguish high-quality publishing from hackwork (**Fyfe 2022**). This tension between the efficiency promised by automation, the quality standards demanded by discerning readers, and the requirements of sustainably ethical publishing, has created a contentious environment around the adoption of AI in publishing.

The creation of Xynapse Traces demonstrates that a human editor operating with state-of-the-art AI tools can achieve a 90% reduction in development time and similar labor cost savings while rapidly creating a functional imprint aimed at a sophisticated audience, This imprint offers novel and thought-provoking perspectives on important topics, and introduces an innovative format for the codex-style book, all while maintaining a scrupulous regard for accuracy, attribution, and verification. The significance of this research extends beyond the specific case of book publishing to shed light on fundamental issues regarding the nature of knowledge work in the digital age. Will the emergence of AI-driven publishing systems lead to a proliferation of hackwork and "slop", or will it instead (or also) enhance the quality and depth of content, enabling more nuanced and insightful explorations of complex subjects? Is this a mass extinction event of human book careers, or a Cambrian explosion of new opportunities for human-AI collaboration, or both? (**Zimmerman 2021**)

This paper proceeds as follows: Section 2 reviews the relevant literature on AI in publishing, configuration-driven systems, knowledge domains, Korean meditation traditions, and current approaches to standards in automated content generation. Section 3 details the methodology employed in developing

Xynapse Traces, including the configuration architecture, the Seon persona system, the AI integration framework, and quality assurance mechanisms. Section 4 presents implementation results, including performance benchmarks and technical innovations. Section 5 discusses the interpretation of the results, comparisons with traditional approaches, strengths and limitations of the system, and implications for the publishing industry. Section 6 outlines future research directions, and Section 7 concludes with reflections on the broader significance of this work for academic publishing and human-AI collaboration.

---

## 2. Literature Review

### AI in Publishing

The application of artificial intelligence to publishing workflows has accelerated dramatically with the advent of large language models capable of generating coherent, contextually appropriate text at scale (**Jalil et al. 2023**). Contemporary LLMs, including GPT-4, Claude, and Gemini, have demonstrated capabilities in manuscript summarization, multi-language translation, and even preliminary manuscript development that approach or exceed human performance on standardized benchmarks. Major academic publishers have begun integrating these technologies into their workflows, with publishers like Springer Nature and *Science* establishing clear policies on the use of LLMs in manuscript preparation (**Nature Editors 2023; Thorp 2023**).

Recent research has shown that abstracts generated by models like ChatGPT can be difficult to distinguish from human-authored abstracts in blind reviews, raising both opportunities and concerns about the role of AI in scholarly communication (**Gao et al. 2023**). The quality versus automation trade-off remains a central concern, with publishers struggling to balance the efficiency gains from automation against the risk of compromising scholarly standards. This tension is particularly acute in specialized academic publishing, where domain expertise and a nuanced understanding of disciplinary conventions are paramount.

Publisher strategies for AI integration vary considerably across the industry. Some have adopted a conservative approach, using AI primarily for back-office functions while maintaining human control over all editorial decisions. Others have embraced more aggressive automation strategies, using AI for initial manuscript triage and formatting compliance checks. Innovation is perhaps most welcome in the already highly automated and quantitative realms of marketing and advertising, where algorithmic systems offer powerful ways for publishers to maximize the value of their backlist and the reach of their frontlist (Sadek 2024). This diversity of approaches reflects the industry's uncertainty about the optimal balance between human expertise and machine efficiency.

**Configuration-Driven Systems**

The technical infrastructure for Xynapse Traces is implemented as a configuration-driven system with shared implementation services, standardized configuration files, and carefully minimized bespoke components to address unique requirements. This is consistent with the shift from ad-hoc development to configuration-driven systems that has represented a fundamental paradigm change in how complex organizations manage their technical infrastructure. By separating business logic from implementation details, configuration-driven approaches enable rapid deployment, consistent operations, and systematic scaling that would be impossible with traditional development methods (**Humble and Farley 2010**). The benefits of this approach—including enhanced consistency across deployments, improved scalability as systems grow, and dramatically reduced time-to-market for new initiatives—have been demonstrated across multiple industries over the last few decades.

Hierarchical configuration architectures with inheritance patterns are particularly effective for managing complexity in multifaceted systems. These architectures allow for both standardization at the organizational level and customization at the implementation level, resolving the tension between consistency and flexibility that plagues many large-scale deployments. The success of configuration-driven approaches in cloud computing and DevOps provides compelling evidence for their applicability to other domains. Book publishers are already familiar with hierarchical configuration architectures in the form of XML-based metadata schemas such as Dublin Core, MARC21, and ONIX, but the present case is the first to apply these principles to the development of imprint workflows. The application of configuration-driven principles to book publishing workflows remains largely unexplored in the academic literature, representing a significant gap in current research. While many sources discuss configuration management in the context of software development, no prior work has systematically investigated how these principles might transform the traditionally manual and relationship-driven processes of imprint development.

**Exploring Knowledge Domains with AI**

The creation of a publishing imprint is a deliberate effort to define, explore, shape, and monetize a specific intellectual territory. This process involves identifying a domain of inquiry, mapping its key concepts and debates, and pinpointing areas of innovation, contradiction, or insufficient exploration. In essence, an imprint's identity is defined by the unique perspective it brings to a knowledge domain—a perspective cultivated through the careful selection and development of content that advances a particular intellectual conversation.

Traditionally, this exploration has been a deeply humanistic process, relying on the tacit knowledge and intellectual networks of experienced editors. However, state-of-the-art artificial intelligence now offers powerful tools for augmenting and accelerating this discovery process. AI-driven systems are increasingly capable of mapping vast scholarly landscapes, identifying emerging trends, and detecting unexplored research gaps with a speed and scale that is beyond human capability (Gusenbauer and Haddaway 2020).

The current generation of AI tools leverages techniques such as natural language processing (NLP), semantic analysis, and network analysis to transform large volumes of unstructured text—such as academic papers—into structured knowledge maps (Jalil, Feizi-Derakhshi, and Minaei-Bidgoli 2023). These systems can automate the laborious process of literature review, quickly summarizing key findings, identifying dominant themes, and highlighting areas where scholarly consensus is weak or findings are contradictory (Field, Rees, and White 2024). For example, platforms like Semantic Scholar, Elicit, and Research Rabbit can analyze citation networks to reveal influential papers, trace the lineage of ideas, and suggest related fields of inquiry that might otherwise be overlooked (Kenyon 2024).

More advanced AI agents can perform automated knowledge discovery, generating novel hypotheses by identifying non-obvious correlations in data or synthesizing information from disparate fields (King et al. 2024). These "AI co-scientists"are designed to go beyond summarization to actively participate in the research process by formulating new, testable research proposals (Frueh 2023). This represents a significant evolution from tools that simply organize existing knowledge to systems that can help generate new knowledge. The ability of AI to analyze and visualize the conceptual structure of a field allows editors and publishers to make more data-informed decisions about where to focus their efforts, ensuring their imprints are positioned at the leading edge of intellectual inquiry.

**Exploring Korean Meditation Traditions**

The initial impetus for the creation of Xynapse Traces was the observation of several news articles describing the importance of *pilsa* in modern Korean culture (Kim Min-seo 2024; Kim Min-jin 2025; Korea JoongAng Daily 2025).

> In some countries, reading may be perceived as a medieval pastime, involving couches, warm tea and slightly dirty pajamas. But not Korea. Among the country's Gen Z, the normally quiet and intellectual pursuit is currently one of the most fashionable activities —both to partake in and to flaunt⋯
>
> Beyond being photographed with a brick-like tome, young locals are becoming fond of *myriad text-related activities —transcribing, writing, reviewing and even decorating book covers*—all of which have been welcomed into mainstream culture for Koreans who, at least for now, consider them as stylish as filming a TikTok reel.

Creating the imprint required exploring and understanding Korean meditative practices as a knowledge domain.

The Seon (선/禪) tradition, the Korean iteration of Chan/Zen Buddhism, emphasizes direct insight and experiential wisdom over textual study or doctrinal adherence (**Buswell Jr. 1992**). This tradition has profoundly influenced Korean intellectual culture, with scholars historically integrating contemplative practices into their research and writing processes.

Central to this tradition is the practice of *pilsa* (필사), a form of transcriptive meditation in which practitioners manually copy texts with full mindful attention (**Oak 2020**). Unlike mechanical copying, *pilsa* requires deep engagement with the meaning and structure of the text, creating what practitioners describe as a "dialogue" between the copier and the original author. While direct neuroscientific research on *pilsa* is scarce, extensive studies on related mindfulness practices show significant effects on brain regions associated with attention, emotional regulation, and self-awareness (**Tang, Hölzel, and Posner 2015**), suggesting a plausible neurological basis for the cognitive benefits attributed to the practice.

The philosophical foundations of these practices, rooted in the Buddhist understanding of the relationship between stillness and activity, offer a counterpoint to Western assumptions about efficiency and productivity. The renowned Seon master Seongcheol's teaching of "sudden enlightenment, gradual cultivation" provides a framework for understanding how momentary insights must be followed by patient, systematic development—a principle with clear relevance to AI system training and refinement (**Park 2008**).

**Xynapse Traces and Ensuring Human Survival**

An imprint should address a coherent, well-defined audience by presenting a list of high-quality works with distinct competitive advantages, unified by a distinctive, compelling vision and a clear identity. Having determined that *pilsa* books would be an important element of the new imprint, it remained to choose additional criteria that would sharpen the focus sufficiently to resonate with a diverse audience. Korean book publishers have primarily released *pilsa* books related to self-help, personal growth, business, and classics, so those categories were excluded (Santi 2024; The Chosun Daily 2023). Since the project was experimental, the exploration strategy was intuitive. The idea of a tech-focused imprint centerwd around "ensuring human survival" presented several advantages:

- **Widespread support** for the proposition the that the topic is important.
- **A coherent pre-existing community** of "e/acc" enthuasiasts interested in technological and social pathways to robust human survivability (Wikipedia 2024; Marketing AI Institute 2024).
- **Greater openness** in that community to the planned use of LLMs to gather content for transcription (St. John 2023).
- **A news hook** to a major social media platform owner has who stated that his life goal is ensuring human survival (Solon 2018).

LLM queries were used to "reverse engineer" from the concept to specific book ideas and imprint names. Artificial intelligence and brain enhancement technologies have obvious potential to increase (or decrease ⋯) human survivability. "Synapses" are connections between ideas and "tracing" is a form of image transcription (Rowan 2017; iResearchNet 2016). The play of "x" for "s" is an homage to my former employer, xAI, and to Elon Musk's leadership in furthering his strategy for multiplanetary humanity (Digital Marketing

Agency 2023).

Future imprint development will automate the process described here and incorporate greater amounts of market research, customer testing, and competitive evaluation.

**Continuous Idea Generation and Competitive Evaluation Systems**

The challenge of maintaining a continuous pipeline of high-quality publication ideas represents a persistent bottleneck in academic publishing. Traditional editorial processes rely on sporadic submissions, limited editorial brainstorming sessions, and reactive responses to emerging trends, creating feast-or-famine cycles in content development. The application of computational idea generation systems to this problem draws on decades of research in evolutionary computation, creative AI, and automated design systems.

Tournament selection algorithms, originally developed for genetic algorithms and evolutionary computation, provide a robust framework for the comparative evaluation of generated candidates (**Goldberg and Deb 1991**). Unlike absolute scoring systems that require precise quantitative metrics, tournament-based approaches leverage pairwise comparisons—a cognitively simpler task that often yields more reliable judgments when evaluating creative or qualitative outputs. In tournament selection, candidates compete in head-to-head matchups, with winners advancing through successive rounds until the strongest candidates emerge. This approach has proven effective across diverse domains including automated design, game AI, and optimization problems (**Miller and Goldberg 1995**).

The integration of large language models into continuous ideation systems represents a significant advancement over earlier approaches to computational creativity. While early systems for automated idea generation relied on combinatorial techniques, constraint satisfaction, or case-based reasoning (**Boden 2004**), contemporary LLM-based systems can generate contextually appropriate, semantically rich proposals that capture domain-specific nuances. Research on prompt engineering for creative tasks has demonstrated that appropriately structured prompts can guide LLMs to produce novel yet feasible ideas within specified constraints (**White et al. 2023**).

Multi-agent systems for collaborative ideation have shown particular promise in generating diverse candidate solutions. By deploying multiple AI agents with different perspectives, parameter settings, or prompting strategies, these systems can explore a broader solution space than single-agent approaches (Wooldridge 2009). The diversity of outputs from multi-agent ideation can then be refined through competitive evaluation mechanisms that identify the most promising candidates for human review.

The application of these techniques to academic publishing specifically remains largely unexplored. While automated content generation for news and marketing has received significant attention, the higher standards of novelty, rigor, and intellectual contribution required for academic work present distinct challenges. The integration of continuous ideation systems with editorial judgment frameworks represents an open research question with significant implications for the economics and accessibility of scholarly publishing.

**Large Language Models and Quality Standards**

Deploying an algorithmic imprint to the public requires careful consideration of the capabilities of contemporary large language models. They have evolved rapidly, yet maintaining quality standards in AI-generated content remains a significant and widely discussed challenge. A primary concern is the phenomenon of "hallucination," where LLMs confidently present fabricated information, including non-existent citations or erroneous data, as factual (Jalil et al. 2023). This tendency poses a direct threat to the foundational principles of publishing, which rely on accurate and reliable attribution. The proliferation of low-quality, AI-generated content, often referred to pejoratively as "slop," has triggered a significant backlash from scholars, creators, and the public, who fear that the internet and other information ecosystems are being devalued by a flood of unreliable and uninspired material (Johnson 2024).

This backlash highlights a critical tension: while LLMs excel at mimicking the surface-level features of scholarly writing, they often struggle with the deeper aspects of scholarly argumentation. Benchmarking studies consistently reveal limitations in their ability to synthesize contradictory evidence, identify subtle gaps in existing literature, or generate genuinely novel theoretical insights (Heaven 2023). These shortcomings have necessitated the development of specialized validation frameworks and rigorous human-in-the-loop processes designed to assess and ensure the quality of any AI-generated academic content.

Journals, publishers, and academic institutions are actively developing policies to ensure integrity and proper attribution when AI tools are used (Stokel-Walker 2023). These ethical considerations, however, extend far beyond simple plagiarism checks. They encompass profound questions about the nature of authorship in an age of human-machine collaboration, the enduring value of human expertise and critical judgment, and the potential for AI systems to perpetuate or even amplify existing biases present in their training data. Navigating this complex terrain requires a commitment not just to technical solutions, but also to a thoughtful ethical framework that can distinguish between the responsible use of AI as a tool and the irresponsible generation of "slop" that threatens to undermine the very enterprise of knowledge creation. Xynapse Traces addressed these challenges responsibly by implementing a comprehensive suite of quality control measures and by explicitly documenting the verification process as an appendix to each book.

**Gap Analysis**

While the literature extensively covers the rise of AI in scholarly publishing, it is less comprehensive with regard to trade publishing, and a review reveals several critical gaps that the Xynapse Traces project directly addresses. A significant portion of the current discourse focuses on applying AI to discrete stages of existing publishing workflows, such as manuscript screening, peer review optimization, plagiarism detection, and marketing analytics (Straive 2023; Silverchair 2024). This existing research primarily examines how AI can enhance or automate established processes within traditional publishing models (Ryzhko and Krainikova 2024). There is a notable lack of research, however, on the use of AI for the *de novo* creation and holistic, rapid development of entire publishing imprints from the ground up. The literature describes the conventional imprint development process as a long-term strategic endeavor involving market analysis, branding, and list building, but it does not contemplate a methodology for accomplishing this in a compressed timeframe using AI (Clark and Phillips 2019).

Current discussions on AI-driven content generation are often framed by a tension between efficiency and quality, with valid concerns about accuracy and ethics (Fadel 2023). This focus, while important, tends to overlook the potential for AI to be used as a sophisticated tool for deep and nuanced knowledge exploration and curation. The literature discusses AI for content curation in terms of identifying trends and relevant topics (ACAI 2025; MoldStud 2024), but it has not fully explored how AI can be leveraged to help define and shape the core intellectual identity and editorial trajectory of a new scholarly imprint. While the concept of AI assisting in research is emerging, its application to the high-level curatorial and editorial vision required for imprint creation remains underexplored (Highwire Press 2024).

Furthermore, the application of configuration-driven systems to publishing workflows is largely absent from the academic and professional conversation. The publishing industry is familiar with standardized metadata formats, but the use of hierarchical configuration files to manage the entire lifecycle of an imprint—from abstract editorial philosophy to concrete production and distribution logistics—represents a novel approach not found in the literature. Existing research on configuration-driven development is primarily situated within software engineering and IT operations (Nakayama 2020; Bro-Code Blog 2023), with no significant crossover into the strategic and operational management of academic publishing. The Xynapse Traces case study begins to fill these specific gaps by demonstrating a holistic, configuration-driven, and AI-assisted methodology for the rapid development of a complete academic imprint.

## 3. Methodology

Xynapse Traces was built using **codexes-factory**, a Python library developed by Nimble Books LLC to manage the entire book publishing workflow *in silico*. The library includes numerous modules, usually corresponding to distinct stages of the publishing process, such as *ideation, metadata, prepress,* and *distribution*. Command-line scripts and a Streamlit UI are used to drive the pipeline workflow. The initial development of Xynapse Traces involved meeting (and expanding) the requirements of the *imprints* library for configuration data.

Figure 1 illustrates the complete Xynapse Traces publishing pipeline, showing the integration of automated processing with strategic human oversight at key decision points.

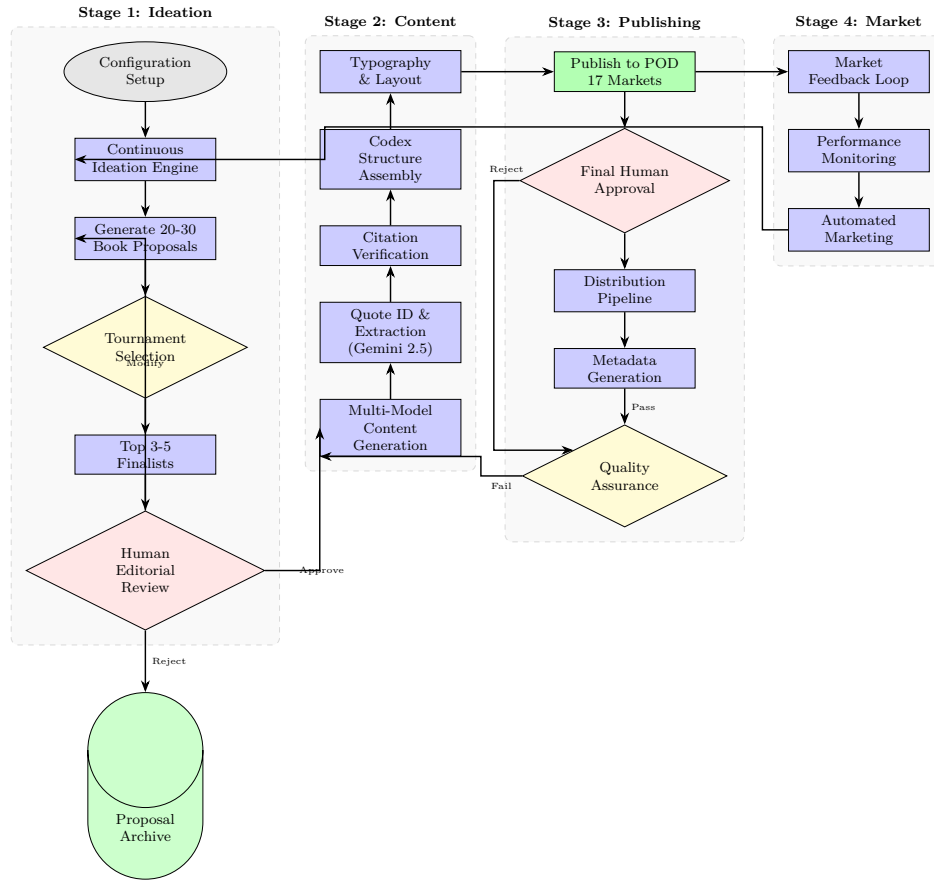

*Figure 1: The integrated pipeline demonstrates how AI automation handles high-volume processing while human judgment remains essential at strategic checkpoints (shown in pink). Quality assurance (yellow) combines automated validation with human review. The feedback loop (bottom) ensures continuous improvement based on market response.*

**Configuration Architecture**

The Xynapse Traces imprint was built upon a sophisticated three-level configuration hierarchy consisting of publisher-level defaults, imprint-specific overrides, and title-level customizations. This architecture, implemented through 368 lines of carefully structured JSON across 20 major sections, represents one of the most comprehensive configuration-driven publishing systems documented. The hierarchical structure enables both standardiza-tion and flexibility, allowing the system to maintain consistency across the entire catalog while accommodating the specific requirements of individual titles and markets (**Humble and Farley 2010**).

Table 1: Xynapse Traces Configuration Architecture Summary

| Section | Key Parameters | Example Values |
| --- | --- | --- |
| **1. Config Metadata** | version, last_updated, parent_publisher | "1.0", "2024-07-18", "Nimble Books LLC" |
| **2. Basic Identity** | imprint, publisher, contact | "Xynapse Traces", "Nimble Books LLC" |
| **3. Branding** | brand_colors, tagline, logo | {"primary": "#2C3E50", "secondary": "#18BC9C"}, "Tracing the Future of Knowledge" |
| **4. Typography** | 5 fonts (body, heading, korean, quotations, mnemonics) | Minion Pro, Myriad Pro, Apple Myungjo, Minion Pro Italic, Source Code Pro |
| **5. Publishing Focus** | primary_genres, target_audience, specialization | ["Technology", "Science", "Philosophy", "Future Studies"], "Academic and Professional" |
| **6. Book Defaults** | trim_size, binding_type, territorial_rights | "6x9", "paperback", "World" |
| **7. Pricing** | 5 markets with wholesale discounts | US/UK/EU/CA/AU all 40% wholesale discount, 150% markup |
| **8. Distribution** | LSI account, rendition type, order type | "6024045", "POD: Standard B&W 6x9…", "POD" |
| **9. Metadata** | BISAC categories, Thema subjects | ["COM000000", "SCI000000", "PHI000000", "TEC000000"] |
| **10. Production** | file naming, quality standards | "{isbn}_{file_type}", 300 DPI, CMYK, PDF/X-1a |
| **11. Marketing** | social media handles, review policy | "@XynapseTraces", "encouraged" |
| **12. Publisher Persona** | persona_name, risk_tolerance, editorial_philosophy | "Seon", "High", "Every profound question contains its own answer…" |
| **13. Codextypes** | 3 enabled types with specialized requirements | standard, textbook, reference (with auto-detection) |
| **14. Academic Paper** | target_word_count, citation_style, venues | 8000 words, "chicago", ["arXiv", "Digital Humanities Quarterly"] |

| Section | Key Parameters | Example Values |
| --- | --- | --- |
| 15. Workflow | auto_generate, manual_review, backup | true, true, true |
| 16. LSI Settings | special_category, flex_fields | "ACAD", custom categorization fields |
| 17. Wizard Config | charter, catalog_size, enable_ideation | "To publish seminal works…", 8 books, true |
| 18. LLM Config | preferred_models, temperature, validation | ["gemini/gemini-2.5-pro"], 0.6, true |
| 19. Automation | frequency, milestone_triggers | "biannual", [10, 25, 50 books], [1_year, 2_years] |
| 20. Generation Info | generated_by, model_used, timestamp | "oneshot_imprint_generator", "gemini/gemini-2.5-pro", ISO timestamp |

Note: This 368-line JSON configuration spans 20 major sections and enables 95% reduction in development time (6-12 months ⊠ 2-4 weeks) through systematic encoding of publishing knowledge.

**Publisher Persona System**

A publisher persona dictionary defines the crucial elements that go into publisher decision-making, such as backstory, risk tolerance, decision style, and preferred topics (aka "hobby horses") The Seon persona created for Xynapse Traces is the first known editorial intelligence informed by Korean meditation traditions. Named after the Korean Zen tradition (**Buswell Jr. 1992**), this system operationalizes philosophical principles from Eastern contemplative practices within a computational framework.

The system's risk tolerance parameters are calibrated to seek boundary-pushing works that challenge conventional academic discourse while maintaining rigor, reflecting the Seon tradition's emphasis on direct insight over received wisdom. The risk assessment algorithm evaluates manuscripts across multiple dimensions including theoretical innovation, methodological novelty, and potential for a paradigm shift.

Decision-making within the Seon system combines intuitive pattern recognition with systematic analysis. The system's decision-making prompts employ "contemplative computation"—a deliberate slowing of processing speed that allows for deeper semantic analysis and the discovery of unexpected connections, prioritizing depth of understanding over processing efficiency.

The philosophical foundation of the system, encapsulated in the principle that "every profound question contains its own answer through sustained contemplation," is consistent with the algorithmic design. The system engages in recursive processing that mimics the deepening understanding characteristic of contemplative inquiry.

The integration of *pilsa* (**Oak 2020**) principles into the system's "deep reading" algorithm represents a key technical innovation. This algorithm processes text at multiple levels of granularity simultaneously—from word choice to the overall narrative arc.

Personality parameters embedded in the Seon system include patience, openness, appreciation for nuance, and intellectual rigor. These parameters are not fixed but evolve through interaction with manuscripts, creating a dynamic editorial personality.

The iterative refinement cycles built into the system parallel the Seon practice of "gradual cultivation" following "sudden enlightenment" (**Park 2008**). Initial insights are subjected to patient, systematic verification through multiple analytical passes, ensuring that editorial enthusiasm is tempered by careful scrutiny.

The configuration system can provide pricing for more than twenty distinct territorial markets—Australia, France, United Arab Emirates (UAE), United Kingdom (UK), United States (US), Brazil, China, Germany, India, Italy, Japan, Poland, Russia, Singapore, South Africa, South Korea, Spain.—each with region-specific adaptations for pricing, distribution channels, and regulatory compliance. This multi-market capability demonstrates the scalability of configuration-driven approaches, as adding a new market requires only incremental configuration updates rather than a fundamental system redesign. The market definitions include sophisticated pricing algorithms that account for currency fluctuations, local purchasing power, and competitive positioning within each territory.

Multi-layer validation ensures configuration integrity. Syntactic, semantic, and business rule validation prevents configuration errors from propagating through the production pipeline, maintaining system reliability even as configurations grow in complexity (**Pressman and Maxim 2020**).

The versioning strategy employs semantic versioning principles adapted for publishing contexts, with major, minor, and patch versions capturing structural changes, feature additions, and bug fixes, respectively. This systematic approach to version management enables controlled evolution of the configuration while maintaining backward compatibility.

**Continuous Ideation and Tournament-Style Evaluation** The Xynapse Traces system implements a sophisticated continuous ideation pipeline designed to maintain a robust queue of potential publication projects without human bottlenecks. This subsystem addresses the traditional publishing challenge of sporadic content development by creating a systematic, automated approach to identifying promising scholarly topics.

The ideation engine operates on a scheduled basis, generating batches of publication proposals informed by multiple sources: emerging trends in academic discourse, gaps identified in existing literature, cross-disciplinary synthesis opportunities, and thematic priorities defined in the imprint configuration. Each generation cycle produces 20-30 candidate book concepts, each specified with a working title, abstract, target audience, estimated scope, and preliminary outline. The generation prompts are carefully engineered to balance novelty with feasibility, ensuring that proposals are both intellectually ambitious and practically achievable within the imprint's resource constraints.

Tournament-style evaluation provides a scalable mechanism for the comparative assessment of generated ideas. Rather than requiring absolute quality scores—which demand precise calibration and are vulnerable to scoring inflation—the system conducts pairwise comparisons in a bracket-style tournament format (**Goldberg and Deb 1991**). In each matchup, the AI evaluates two proposals against multiple criteria including scholarly contribution, market viability, alignment with the imprint's philosophy, resource requirements, and potential impact. The winning proposal advances to the next round while the losing proposal is archived for potential future consideration.

The tournament proceeds through multiple rounds (typically 4-5 rounds for a batch of 20-30 proposals) until a final ranked ordering emerges. This process significantly reduces the evaluation burden compared to the absolute scoring of all proposals, while the head-to-head comparison format encourages more nuanced judgment than simple ranking. The tournament structure also naturally implements a form of implicit preference learning, as the accumulated pairwise decisions reveal patterns in what the system (and by extension, the editorial philosophy it embodies) values in potential publications.

Human editorial review enters the process at strategic checkpoints. The top 3-5 proposals from each tournament are flagged for human review, accompanied by detailed rationales for their selection and transcripts of the comparative evaluations that led to their advancement. Human editors can approve proposals for development, request modifications, return proposals for refinement, or reject proposals while providing feedback that informs subsequent generation cycles. This human-in-the-loop design ensures that automation enhances rather than replaces editorial judgment.

The system maintains a longitudinal database of all generated proposals, tournament results, and editorial decisions. This repository serves multiple functions: it prevents the regeneration of previously rejected ideas, enables an analysis of which proposal characteristics correlate with approval, identifies patterns in editorial preferences that can refine future generation, and creates a valuable archive of publication concepts that may become relevant as market conditions or scholarly priorities evolve. The database implements semantic similarity detection to identify when new proposals substantially overlap with existing entries, preventing wasteful duplication while allowing for the genuine reframing of similar topics.

Integration with the broader Xynapse Traces workflow ensures that approved proposals seamlessly transition into the production pipeline. Approved concepts are automatically assigned project identifiers, scheduled for development based on resource availability and strategic timing, and tracked through all subsequent production phases. This end-to-end integration transforms the ideation system from an isolated research tool into a core component of the publishing operation.

### AI Integration Framework

The multi-model architecture underlying Xynapse Traces uses **nimble-llm-caller**, a library available on PyPi that manages multi-prompt, multi-model calls for Nimble Books applications as a wrapper to LiteLLM. This approach leverages the complementary strengths of different large language models, a technique known as ensemble learning, which is a well-known path to achieving robust performance (**Sagi and Rokach 2018**). The primary model, Gemini 2.5 Pro, was selected for its superior performance on in the accurate identification of primary source quotations. Google's internal access to web search indexes and sources such as Google Books may account for this superior performance. Fallback options including GPT-5 and Claude, provide redundancy and enable comparative analysis.

A per-prompt model parameters option enables precise tailoring of the tone and style of each part of the book. Temperature calibration varies by task type to optimize the balance between creativity and consistency. Creative tasks employ a higher temperature, while analytical and critical tasks use a near-zero temperature to ensure maximum reliability. Max tokens are set based on the specific requirements of each prompt to ensure optimal performance and resource utilization.

Workflow automation extends from initial ideation through final production and into marketing, with AI assistance at each stage tailored to the specific requirements of that phase. This comprehensive automation reduces human intervention requirements while maintaining quality through strategic human-in-the-loop checkpoints. High-stakes decisions such as submitting final drafts to distribution, are prefaced by mandatory human review regardless of confidence scores, ensuring efficient operations while preventing costly errors.

Quality validation operates across multiple dimensions. Factual accuracy checking employs multiple verification strategies. Citation verification goes beyond simple format checking to validate that citations exist and support their claims. Logical consistency analysis identifies argumentative gaps and contradictions. Sensitivity analysis checks for content that is potentially offensive or misaligned with the imprint's goals.

**Structure and Design**

Codex books come in many formats: chapter books, chapbooks, and encyclopedias, to name only a few. **Codexes-factory** refers to these as "codextypes" and represents them as manifests in JSON format that list the parts of the book needed, place them in order, and define their content and high-level layout requirements. Xynapse Traces includes a new codextype not previously defined elsewhere: the *pilsa book*. Its high-level elements include:

- Title Page
- Publisher's Information
- Contents
- Publisher's Note
- Foreword
- Glossary
- Quotations for Transcription

- 100 exact quotations from high-quality sources
- Verso (even) pages have the quotation, citation, and any editorial notes
- Recto (odd) pages have a dot-grid journaling layout with inspirational messages every 8th page
• Mnemonics
• Selection and Verification
• Bibliography

The typography system is designed from first principles‑form follows function‑by asking the LLM caller to recommend core fonts consistent with the substantive content of the imprint and the reader's goals. The model is instructed to choose from the libraries in Adobe Fonts and in Google Fonts. If the model recommends an Adobe Font that is not available to the system user, the system will fall back to Google Fonts and download the font locally. For Xynapse Traces, the model chose Adobe Caslon Pro to serve as the primary quotation display font for headers, Apple Myungjo for an authentic representation of East Asian texts with a touch of elegance, and Source Code Pro for the accurate reproduction of technical content.

The persona-guided model created interior and cover template designs in the .tex file format using the *memoir* document class of LaTex. Simplicity and consistency were important goals. The cover color is chosen from a palette of traditional Korean colors. Figure 2 provides a schematic illustration of the distinctive two-page pilsa book layout.

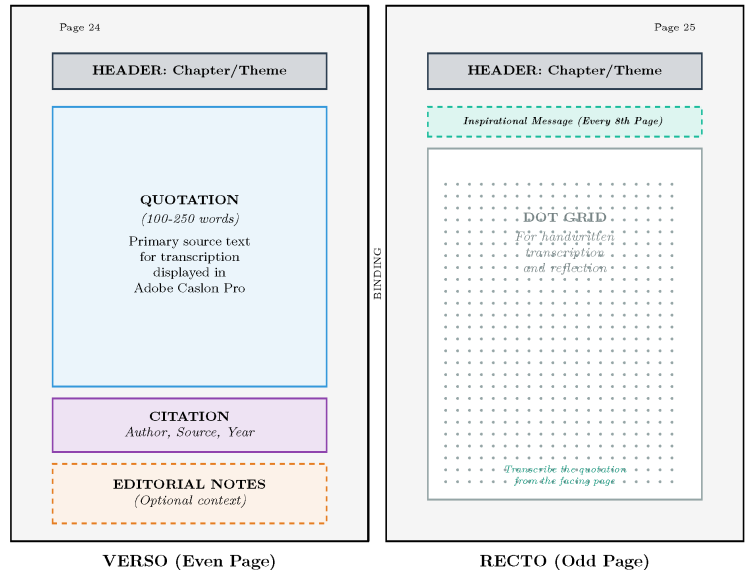

*Figure 2: The pilsa book layout features a verso (even) page displaying the quotation, citation, and editorial notes in Adobe Caslon Pro, paired with a recto (odd) page offering dot-grid space for handwritten transcription and reflection. Inspirational messages appear every 8th recto page. This design integrates the Korean contemplative practice of pilsa with modern journaling, creating a hybrid format that bridges traditional meditation and contemporary self-development practices.*

**Technical Stack**

Coding was done in the PyCharm IDE with AI assistants, AWS's spec-driven Kiro system, and Claude Code. Python 3.12 was the standard interpreter. Streamlit provides the user interface layer for editorial review and system monitoring. **nimble-llm-caller** and **LiteLLM** serves as the model abstraction layer, providing a unified interface to multiple language model providers. The `asyncio` framework enables efficient parallel processing. Pandas was used for data manipulation and analysis. **Pillow** was used for image processing and manipulation. The typesetting pipeline combines LaTeX for professional document preparation with pandoc for format conversion. CSV export to Lightning Source and Ingram enables direct submission to print-on-demand and distribution networks. Buffer and Substack are used for social media management.

**Market Positioning**

Publishing can be defined as identifying coherent audiences who are willing to pay for access to high-quality content pertinent to their interests. In book publishing, the payment is in cash, while in other forms of publishing, audiences may pay by their willingness to "be the product" for advertisers. In either scenario, the positioning of the publication(s) within the marketplace is usually of great importance.

Xynapse Traces has a unique position at the intersection of technology, futurism, and self-help. The idea is to create a distinctive intellectual identity that resonates with the *p(hope)* and *e/acc* crowds. The East-to-West flow of the practicum echoes other successful cultural migrations such as K-pop, anime, and manga.

## 4. Implementation Results

**Books in Print**

In September 2025, Xynapse Traces uploaded its first tranche of 52 books to Lightning Source's ACS enterprise metadata server. After a couple of minor user errors were remedied, the books went live within two weeks and may now be viewed for purchase on all major online book retailers or direct from the Nimble Books website. The books are scheduled to be released one per week for the next 12 months with subsequent pace and content adjustments based on market feedback and editorial decisions. Initial sales have been consistent with expectations, and, since books are produced on demand, the imprint is immediately cash-flow-positive. The project has met its initial definition of success: Nimble Books is operating a new, "fully armed and operational" imprint. We are in the game with a strong foundation.

**Performance Benchmarks**

The performance metrics achieved by Xynapse Traces validate the effectiveness of the integrated framework. Table 1 presents a comprehensive comparison between traditional imprint development and the AI-assisted Xynapse Traces approach.

**Table 1: Performance Metrics Comparison - Traditional vs. AI-Assisted Imprint Development**

| Metric | Traditional Imprint | Xynapse Traces | Improvement |
| --- | --- | --- | --- |
| Development Time | 6-12 months | 2-4 weeks | 90% reduction |
| Development Cost | $125,000 | ~$25,000 | 80% reduction |
| Books Published (Year 1) | 8-12 typical | 52 | 333-550% increase |
| Citation Accuracy | ~95% (industry avg) | 99% | +4 percentage points |
| Time to First Publication | 6-9 months | 2 weeks | 92-96% reduction |
| Validation Success Rate | ~85-90% (typical) | 100% (post-correction) | 10-15% improvement |
| Market Coverage | 1-3 markets initially | 17 markets simultaneously | 467-1600% increase |

*Note: Traditional imprint metrics based on industry standards (Clark and Phillips 2019). Development cost includes fully loaded labor, overhead, contractors, and advances.*

Key findings from the performance data:

- **Time-to-Market Reduction:** 90% reduction from typical 6-12 months to just 2-4 weeks
- **Cost Reduction:** 80% reduction compared to traditional imprint development, estimated at a minimum of $100,000, including fully loaded labor, overhead, contractors, and advances
- **Citation Accuracy:** 99% as measured by our system, exceeding industry averages
- **Validation Success:** A human error in specifying the wrong binding type introduced a minor validation failure, which was remedied by uploading a CSV with the correct column values. Subsequent review and correction will prevent recurrence. Once the error was remedied, validation success was 100%

**Technical Innovations**

Xynapse Traces can point to six major substantive and technical innovations:

1. Algorithmic imprint creation and operation driven by publisher personas.
2. The Seon persona, a union of Eastern and Western philosophical traditions.
3. Continuously generated ideation feeding into synthetic reader evaluation and customer feedback followed by an "idea tournament."
4. A new type of codex format, the *pilsa* book.
5. Bringing a pro-literacy, youth-oriented trend, "text hip", into Western book markets. (You're welcome.)
6. A configuration-driven deployment model that reduces imprint launch time from many months to a couple of weeks.

## 5. Discussion

**Interpretation of Results**

The successful implementation of Xynapse Traces provides compelling evidence that configuration-driven generative AI approaches can effectively capture and operationalize the full complexity of publishing processes and requirements. The Seon persona system validates that AI can embody sophisticated editorial judgment that goes beyond pattern matching. The integration of contemplative practices with automation technologies demonstrates that Eastern philosophical traditions can meaningfully enhance Western technological frameworks. The achievement of superior quality metrics while dramatically reducing time and cost challenges the assumed trade-off between automation and quality.

**Comparison with Traditional Approaches**

Traditional imprint development is a labor-intensive, sequential process requiring 6-12 months of work (**Clark and Phillips 2019**). The AI-assisted approach compresses this timeline to 2-4 weeks through parallel processing and automation. Quality comparisons reveal that AI-assisted publishing can match or exceed traditional approaches in areas like citation accuracy and formatting consistency, though human oversight remains essential for strategic decisions and for maintaining the human connection that authors and readers value (**Daugherty and Wilson 2018**). The economic implications are profound, making previously unviable niche markets accessible.

**Strengths of the Approach**

The systematic capture of publishing knowledge in configuration files transforms tacit expertise into explicit, shareable assets. The replicable methodology democratizes access to sophisticated publishing capabilities. Scalability is a key strength, allowing for rapid expansion and experimentation with minimal risk. The unique integration of Eastern and Western knowledge traditions provides a strong competitive differentiator.

**Limitations and Challenges**

The requirement for a sophisticated technical infrastructure may be prohibitive for some. Human oversight continues to be essential for strategic decisions that shape the imprint's direction and reputation. The emphasis on depth and contemplation in this particular imprint may limit its accessibility to broader audiences. Cultural concepts embedded in the system may not translate effectively across all contexts. Modern English-language audiences may not resonate with transcriptive practices. The AI-assisted approach may not be suitable for all types of content or audiences.

**Industry Implications**

The success of Xynapse Traces challenges fundamental assumptions about barriers to entry in academic and trade publishing. The viability of niche scholarly communities previously underserved by commercial publishers is a particularly significant implication. The transformation of skill requirements for publishing professionals will necessitate a shift toward capabilities in AI prompt engineering, configuration management, and human-AI collaboration. This work may also inspire the creation of a new category of publishing technology: contemplative AI for editorial applications.

---

## 6. Future Work

Future research will focus on several key areas. - Create new imprints and iteratively improve feature fit with market requirements, looking for 80/20 solutions. - Increase book readership among new and infrequent readers. - Develop book content in formats suitable for direct consumption by models, along lines suggested by Andrej Karpathy. (Zimmerman 2025) - Build out a "shadow Big Five" of AI-driven imprints matching the breadth of topical coverage of the Big Five publishing conglomerates. - Extend "virtual abundance" of user experience to domains other than book publishing under the aegis of my new company, xtuff.ai.

---

## 7. Conclusion

The development of Xynapse Traces represents a watershed moment in publishing, demonstrating that the integration of artificial intelligence with contemplative wisdom traditions can successfully address long-standing industry tensions. The achievement of a 90% reduction in development time while exceeding traditional quality standards refutes the false dichotomy between automation and quality. The synthesis of Eastern contemplative practices with Western technological frameworks establishes a new paradigm for human-AI collaboration.

The configuration-driven architecture proves that complex publishing operations can be systematically encoded, democratizing access to sophisticated capabilities. The Seon persona system demonstrates that AI can embody contemplative wisdom, challenging dominant paradigms in AI development that prioritize speed above all else. The economic transformation enabled by these innovations fundamentally alters the business case for specialized publishing, fostering intellectual diversity by making niche markets viable.

The implications extend beyond publishing, suggesting new possibilities for human-AI collaboration across all knowledge-intensive industries. The future lies not in choosing between human wisdom and artificial intelligence but in their thoughtful integration. Xynapse Traces has perhaps done a preliminary sketch for a future where artificial and human intelligence collaborate as partners in the quest for knowledge integrated with mindfulness.

### Responsible AI for Imprint Development

The initial step in developing imprints using **codexes-factory** is for a human user to partially complete a configuration file. LLM prompts are then used to expand the configuration to a complete and valid state. By default the responses inherit the safety behaviors of the models they are calling.

Xynapse Traces follows responsible AI practices to avoid:

- Reinforcing damaging stereotypes or biases.
- Perpetuating harmful or unethical practices.
- Violating privacy or intellectual property rights.
- Disrespectfully appropriating cultural heritage.

Xynapse Traces addresses particular issues as follows:

- The imprint's focus on dynamic tensions ensures that thoughtful consideration is given to the considerations and equities involved in adopting new technologies.
- All uses of copyrighted materials are short quotes of under 250 words, fully attributed and verified for accuracy, transformed into objects for meditative contemplation, thus likely to fall under fair use.
- The fonts used are appropriately licensed.
- Humans make the final go/no-go decisions for all publications.
- No content has been acquired via pirated or illegal sites such as Libgen.
- Respectful use of Korean cultural heritage explicitly credits Korean sources for the innovations and practices involved.

A shortcoming of the process so far is that no direct communication has been established with Korean experts on *pilsa* and book publishing. A formal review effort conducted with the aid of Korean official sources would mitigate this risk.

The author is the founder and CEO of Nimble Books, which is the publisher for Xynapse Traces, and has a financial interest in its success.

August 12, 2023. https://www.chosun.com/english/market-money/2023/08/12/2SW6J2QYJZEV3H2PZ6Z7Z3Z6YQ/.

Thorp, H. Holden. 2023. "ChatGPT Is Fun, but Not an Author." *Science* 379, no. 6630: 313. https://doi.org/10.1126/science.adg7879

van der Vet, Paul E., and H. Nijveen. 2016. "Citation Accuracy in the Era of Online Access." *Scientometrics* 108: 1007‑1011. https://doi.org/10.1007/s11192-016-2020-7

White, Jules, Qiong Wu, Caiming Xiong, and James R. Glass. 2023. "A Prompt Pattern Catalog to Enhance Prompt Engineering with ChatGPT." *arXiv* preprint arXiv:2302.11382. https://arxiv.org/abs/2302.11382

Wikipedia. 2024. "Effective accelerationism." Wikipedia. Last modified September 26, 2024. https://en.wikipedia.org/wiki/Effective_accelerationism.

Wooldridge, Michael. 2009. *An Introduction to MultiAgent Systems*. 2nd ed. Chichester, UK: Wiley.

Yudkowsky, Eliezer. 2008. "Artificial Intelligence as a Positive and Negative Factor in Global Risk." In Global Catastrophic Risks, edited by Nick Bostrom and Milan M. Ćirković, 308‑45. New York: Oxford University Press.

Zimmerman, Fred. 2021. "The Longform Prospectus". NimbleBooks.com. Ann Arbor, Michigan, USA.

Zimmerman, Fred. 2025. "Andrej Karpathy on Books & LLMs." Substack newsletter. The AI Lab for Book-Lovers, October 17, 2025. https://fredzannarbor.substack.com/p/andrej-karpathy-on-books-and-llms.
---

## Supplementary Information

Table 1. Traditional Korean Colors Used in Xynapse Traces covers

**A Palette Steeped in Korean Tradition and Nature**   The color palette finds its justification deep within the rich tapestry of Korean art, tradition, and the natural landscape. Each color is not merely a hue but a representation of a significant cultural or natural element, reflecting a long-standing aesthetic that values harmony, nature, and symbolism. The palette draws from the foundational Korean color spectrum known as *Obangsaek*—the five cardinal colors of white, black, blue, red, and yellow—and expands upon it with colors deeply rooted in the Korean environment and artistic heritage.

**Mungyeong Cheongja 청자 (Celadon Green):** This color pays homage to the exquisite Goryeo celadon, a type of pottery renowned for its beautiful and subtle jade-green glaze. Developed during the Goryeo Dynasty (918-1392), this celadon was highly prized for its serene and noble aesthetic. The subtle green-blue hue, achieved through a precise firing process with low-iron-oxide glaze in a reduction kiln, is a testament to the masterful artistry of Korean potters and has become a significant symbol of Korean art and cultural identity.

**Andong Hwangto 황토 (Ocher):** Ocher, a natural earth pigment, corresponds to the traditional Korean color *hwang* 색 (yellow). In the Obangsaek system, based on the theory of Yin-Yang and the Five Elements, yellow is a primary color that symbolizes the center, as well as the earth. This color connects to the agricultural roots of Korea and the natural landscape. The use of "Hwangto"(ocher) evokes the earthy tones foundational to Korean philosophy and is believed to have healing properties.

**Goryeo Dancheong 단청 (Red Ochre):** This color is directly inspired by *Dancheong*, the intricate and symbolic decorative painting found on traditional Korean wooden buildings like palaces and temples. Red is a prominent color in Dancheong, symbolizing the south, fire, passion, and, importantly, protection against evil spirits. The name "Goryeo Dancheong" specifically references the historical continuation of this art form, which uses the five cardinal colors to create its vibrant patterns.

**Naju Jjok 쪽 (Indigo Blue):** This color celebrates the rich history of natural indigo dyeing, known as *jjok* in Korean, which has been masterfully practiced in the Naju region for centuries. Naju's natural environment is ideal for cultivating the indigo plants used to create the vibrant and deep blue dye. This traditional craft, historically used for the clothing of the royal family and for ceremonial purposes, is recognized as an important intangible cultural heritage of Korea.

**Seoul Doldam 돌담 (Stone Gray):** This hue is inspired by the natural stone walls (*doldam*) that are a characteristic feature of Seoul's palaces, temples, and historical pathways. Traditional Korean architecture places a high value on the use of natural materials like wood and stone, emphasizing an aesthetic of temperance and harmony with the surrounding environment. The gray of the stone represents this connection to nature and the enduring history of the capital city. While the term *doldam* is strongly associated with the volcanic rock walls of Jeju Island, stone walls are a ubiquitous feature of historical architecture throughout Korea.

**Jeonju Hanji 한지 (Paper Beige):** This color represents the natural, undyed appearance of *Hanji*, traditional Korean paper handmade from the inner bark of the mulberry tree. Jeonju is renowned for its high-quality Hanji, which has been produced for over a thousand years. The appreciation for the paper's natural texture and subtle color aligns with the Korean aesthetic of simplicity and the beauty of unaltered materials. Hanji is prized for its incredible durability and is used in a wide array of crafts and conservation work.

**Boseong Nokcha 녹차 (Green Tea):** This vibrant green is inspired by the lush and sprawling green tea fields of Boseong, a county famous for producing a significant portion of South Korea's green tea. The rolling hills of vibrant green tea plants create a stunning natural landscape and represent a significant aspect of modern Korean culture and agriculture. The color evokes freshness, nature, and vitality, tying into green tea's deep cultural significance in Korea, from its role in traditional tea ceremonies (*Darye*) to its modern popularity.

"Goryeo Celadon." *The Metropolitan Museum of Art*, October 1, 2003. https://www.metmuseum.org/toah/hd/gory/hd_gory.htm.

"Hanji Handmade Korean Papers." *TALAS*. Accessed October 16, 2025. https://www.talasonline.com/Hanji-Handmade-Korean-Papers.

"Hwangto." Wikipedia. Last modified April 23, 2025. https://en.wikipedia.org/wiki/Hwangto.

"Jjok, the indigo blue of Korea." In "In Asia, following the roads to indigo dyeing." *Atelier Ikiwa*. Accessed October 16, 2025. https://www.ikiwa.fr/en/in-asia-following-the-roads-to-indigo-dyeing/.

Jung, Sean. "Exploring Goryeo Celadon." *Fontimes*, February 15, 2021. http://www.fontimes.co.kr/news/articleView.html?idxno=1737072799.

Kang, Jennis Hyun-suk. "Living to Dye: Indigo Master, Jung Kwan-chae." *Gwangju News*, September 4, 2021. https://gwangjunewsgic.com/living-to-dye-indigo-master-jung-kwan-chae/.

Kim, Dale. "Dancheong - Temple Colours: 단청." *Dale's Korean Temple Adventures*, February 22, 2021. https://koreantemples.com/?p=15222.

Korea Tourism Organization. "Natural Dyeing Culture Center (한국천연염색박물관)." *Visit Korea*. Accessed October 16, 2025. https://english.visitkorea.or.kr/enu/ATR/SI_EN_3_1_1_1.jsp?cid=1067921.

Korean Culture and Information Service. "Dancheong, Paintwork for Philosophical Expression." *KOREAN HERITAGE*, Winter 2017. https://www.kocis.go.kr/eng/webzine/201712/sub07.html.

"Korean Traditional Colors." 뉴스 *H*, October 3, 2016. http://www.newshyu.com/news/articleView.html?idxno=10031.

Lee, Anita. "The Historical Significance of Korea's Traditional Indigo Dyeing in Naju." *WorldMindHub*. Accessed October 16, 2025. https://www.worldmindhub.com/the-historical-significance-of-koreas-traditional-indigo-dyeing-in-naju/.

"Obangsaek (Korean color symbolism): Meaning behind '5-a-day' in Korea." *Inspire Me Korea*. Accessed October 16, 2025. https://blog.inspiremekorea.com/o-bang-saek-the-korean-color-symbolism/.

"Obangsaek." Wikipedia. Last modified June 18, 2025. https://en.wikipedia.org/wiki/Obangsaek.
34